\documentclass[aps,prd,nofootinbib,amsmath,amssymb,showpacs,twocolumn,superscriptaddress,10pt]{revtex4}
\usepackage{graphicx}
\usepackage{dcolumn}
\usepackage{bm}
\usepackage{amssymb}
\usepackage{latexsym}
\usepackage{booktabs}
\usepackage[colorlinks, linkcolor=blue, citecolor=blue, urlcolor=blue]{hyperref}

\newcommand{\be}{\begin{equation}}
\newcommand{\ee}{\end{equation}}
\newcommand{\bq}{\begin{eqnarray}}
\newcommand{\eq}{\end{eqnarray}}

\begin{document}

\title{Running coupling: Does the coupling between dark energy and dark matter change
sign during the cosmological evolution?}

\author{Yun-He Li}
\affiliation{Department of Physics, College of Sciences,
Northeastern University, Shenyang 110819, China}
\author{Xin Zhang}
\email{zhangxin@mail.neu.edu.cn} \affiliation{Department of Physics,
College of Sciences, Northeastern University, Shenyang 110819,
China} \affiliation{Center for High Energy Physics, Peking
University, Beijing 100080, China}

\begin{abstract}

In this paper we put forward a running coupling scenario for
describing the interaction between dark energy and dark matter. The
dark sector interaction in our scenario is free of the assumption
that the interaction term $Q$ is proportional to the Hubble
expansion rate and the energy densities of dark sectors. We only use
a time-variable coupling $b(a)$ (with $a$ the scale factor of the
universe) to characterize the interaction $Q$. We propose a
parametrization form for the running coupling $b(a)=b_0a+b_e(1-a)$
in which the early-time coupling is given by a constant $b_e$, while
today the coupling is given by another constant, $b_0$. For
investigating the feature of the running coupling, we employ three
dark energy models, namely, the cosmological constant model
($w=-1$), the constant $w$ model ($w=w_0$), and the time-dependent
$w$ model ($w(a)=w_0+w_1(1-a)$). We constrain the models with the
current observational data, including the type Ia supernova, the
baryon acoustic oscillation, the cosmic microwave background, the
Hubble expansion rate, and the X-ray gas mass fraction data. The
fitting results indicate that a time-varying vacuum scenario is
favored, in which the coupling $b(z)$ crosses the noninteracting
line ($b=0$) during the cosmological evolution and the sign changes
from negative to positive. The crossing of the noninteracting line
happens at around $z=0.2-0.3$, and the crossing behavior is favored
at about 1$\sigma$ confidence level. Our work implies that we should
pay more attention to the time-varying vacuum model and seriously
consider the phenomenological construction of a sign-changeable or
oscillatory interaction between dark sectors.

\end{abstract}

 \maketitle

%============================= section 1 ===========================================

\section{Introduction}\label{sec1}

Dark energy (DE) and dark matter (DM) are the dominant components in
the current universe, according to the recent astronomical
observations \cite{Riess98,Tegmark04,Spergel03}. They together
account for about 96\% of the critical energy density of the
universe. However, ironically, we have known little about the
natures of DE and DM. Although DM is ``dark'' (nonluminous), its
gravitational property is normal, i.e., it gravitationally behaves
like the usual baryon matter and thus can form structures in the
universe. The property of DE is much more exotic in that it is
gravitationally repulsive and so responsible for the accelerated
expansion of the universe \cite{dereview}.

Since we are ignorant of the natures of both DE and DM, we cannot
ignore such an important possibility that there is some direct,
non-gravitational interaction between DE and DM. Intriguingly, such
a possible interaction between DE and DM plays a crucial role in
helping solve (or, at least alleviate) several important theoretical
problems of DE. For example, it can be used to understand the cosmic
coincidence problem \cite{intde1}, to avoid the cosmic doomsday
brought by phantom \cite{intde2}, and to solve the cosmic age
problem caused by old quasar \cite{intde3} as well. Therefore, it is
very meaningful to seriously study the interaction between dark
sectors. Owing to the lack of the knowledge of micro-origin of the
interaction, one has to first phenomenologically propose an
interacting DE model and then test its theoretical and observational
consequences. So far, lots of phenomenological interacting DE models
have been studied~\cite{intdeOdintsov,intde4}.

When considering the interaction between dark sectors, the
continuity equations for energy densities of DE and DM are of the
form: \be\label{eq1} \dot{\rho}_{de}+3H(1+w_{de})\rho_{de}=-Q, \ee
\be\label{eq2} \dot{\rho}_{dm}+3H\rho_{dm}=Q, \ee where $\rho_{de}$
and $\rho_{dm}$ are the energy densities of DE and DM, respectively,
$H=\dot{a}/a$ is the Hubble parameter, $a$ is the scale factor of
the Friedmann-Robertson-Walker (FRW) universe,
$w_{de}=p_{de}/\rho_{de}$ is the equation of state (EOS) parameter
of DE, a dot denotes the derivative with respect to cosmic time $t$,
and $Q$ denotes the phenomenological interaction term. Several forms
for $Q$ have been put forward and have been fitted with
observations~\cite{intde5}. Of course, all of these models are
phenomenological. Most of them are constructed specifically for
mathematical simplicity --- for example, models in which $Q\propto
H\rho$, where $\rho$ denotes the energy density of the dark sectors,
and usually it has three choices, namely, $\rho=\rho_{dm}$,
$\rho=\rho_{de}$, and $\rho=\rho_{de}+\rho_{dm}$. In addition, there
are also some models~\cite{11Valiviita:2009nu} in which the
assumption of that $Q$ is proportional to the Hubble parameter is
abandoned and thus $Q\propto\rho$. Such models are designed by
consulting the simple models of reheating, of dark matter decay into
radiation, and of curvaton decay --- i.e., where the interaction has
the form of a decay of one species into another, with constant decay
rate. However, it should be stressed that these models are severely
dependent on the man-made choice of the special interaction forms.
In other words, the predictions and observational consequences are
model-dependent. In particular, the abovementioned phenomenological
models exclude the important possibility that the interaction
changes sign during the cosmological evolution. It is of interest to
point out that a sign-changeable or oscillatory form of interaction
is possible, according to the current
observations.%~\cite{12Cai:2009ht}.

Recently, Cai and Su~\cite{12Cai:2009ht}, without choosing a special
phenomenological form of interaction, proposed a novel scheme in
which the whole redshift range is divided into a few bins and the
interaction term $\delta(z)$ (note that here $Q=3H\delta$) is set to
be a constant in each bin, and by fitting the observational data
they found that $\delta(z)$ is likely to cross the noninteracting
($\delta=0$) line. This study is fairly enlightening and inspires us
to open our mind to seriously consider the possibility of that the
interaction between dark sectors changes sign during the
cosmological evolution. If such an observation is conclusive, it is
suggested that more general phenomenological forms of interaction
term should be put forth. However, the work of Cai and
Su~\cite{12Cai:2009ht} seems not sufficient to prove that the
interaction changes its sign, since their conclusion is drawn based
upon the behavior of the best-fitted $\delta(z)$, and the errors of
the fitting results weaken the conclusion to a great degree due to
the fact that the observational data currently available cannot
determine more than two parameters in a piecewise parametrization
approach.

In this paper, our aim is to verify whether the interaction indeed
changes its sign (i.e., crosses the noninteracting line) during the
evolution, by using a different method. We propose a parametrization
form for the interaction term $Q$. In our work, we further abandon
the assumption that $Q$ is proportional to the Hubble expansion rate
$H$. So, the interaction term $Q$ is only characterized by the
coupling $b$: \be\label{eq3} Q(a)=3b(a)H_0\rho_0, \ee where the
dimensionless coupling $b(a)$ is variable with the cosmological
evolution. Note that here the occurrence of the present-day Hubble
parameter $H_0$ and the present-day density of dark sectors
$\rho_0=\rho_{de0}+\rho_{dm0}$ is only for a dimensional reason. By
the way, in the whole work the subscript ``0'' always indicates the
present-day value of the corresponding quantity. From the form of
Eq.~(\ref{eq3}), we see that the evolution of the interaction term
$Q$ is totally described by the running of the coupling constant,
$b(a)$, so our scenario can be called the ``running coupling''.
Furthermore, we assume that the coupling $b$ is described by a
constant $b_0$ at the late times, and determined by another constant
$b_e$ at the early times; and the whole evolution of $b(a)$ is
totally characterized by the two parameters, $b_0$ and $b_e$. For
continuously connecting the early-time and late-time behaviors, we
put forward the following two-parameter form for the coupling
$b(a)$: \be\label{eqb} b(a)=b_0a+b_e(1-a). \ee Though the
interaction term in our work depends on a particular parametrization
form, the parameters can be tightly constrained by the current
observational data, overcoming the disadvantage of the piecewise
fitting method. The reconstructed evolution of $b(a)$ will indicate
whether the coupling between the dark sectors crosses the
noninteracting line.

In this paper, we will investigate our coupling parametrization with
the latest observational data. For the EOS of DE, $w$, we consider
the following three cases: (1) the cosmological constant (vacuum
energy), $w=-1$; (2) the constant EOS, $w=w_0$; (3) the
time-variable EOS, namely, the Chevallier-Polarski-Linder (CPL)
parametrization, $w(a)=w_0+w_1(1-a)$~\cite{13CPL}. We will fit the
three interacting DE models with the data from the Union2 type Ia
supernovae (SNIa), the baryon acoustic oscillation (BAO), the cosmic
microwave background (CMB), the Hubble expansion rate, and the X-ray
gas mass fraction. We obtain the best-fitted parameters and
likelihoods by using the Monte Carlo Markov chain (MCMC) method. We
will show that the interaction $Q$ between DE and DM indeed changes
sign around $z=0.2-0.3$ during the cosmological evolution, at about
1$\sigma$ confidence level (CL).

%============================= section 2 ===========================================

\section{Methodology}\label{sec2}

We consider interacting DE models in a spatially flat FRW universe.
The Friedmann equation reads \be\label{eq4}
3M_{Pl}^2H^2=\rho_r+\rho_b+\rho_{de}+\rho_{dm}, \ee where $\rho_r$,
$\rho_b$, $\rho_{de}$ and $\rho_{dm}$ are the energy densities of
radiation, baryon, DE and DM, respectively, and $M_{Pl}$ is the
reduced Planck mass. It is convenient to introduce the fractional
energy densities $\Omega_i\equiv\rho_i/3M_{Pl}^2H^2$, with $i =r,\
b,\ de,$ and $dm$. Obviously, \be\label{eq5}
\Omega_r+\Omega_b+\Omega_{de}+\Omega_{dm}=1. \ee

Substituting Eqs.~(\ref{eq3}) and (\ref{eqb}) into Eqs.~(\ref{eq1})
and (\ref{eq2}), and defining the functions
$f_{de}=\rho_{de}/\rho_{de0}$ and $f_{dm}=\rho_{dm}/\rho_{dm0}$, we
obtain \be\label{eq8}
\frac{df_{de}(x)}{dx}+3(1+w)f_{de}(x)=-\frac{3}{E(x)}\left(1+\frac{1}{f_0}\right)\left[b_0e^x+b_e(1-e^x)\right],
\ee \be\label{eq9}
\frac{df_{dm}(x)}{dx}+3f_{dm}(x)=\frac{3}{E(x)}(1+f_0)\left[b_0e^x+b_e(1-e^x)\right],
\ee where $x\equiv\ln{a}$,
$f_0\equiv\rho_{de0}/\rho_{dm0}=\Omega_{de0}/\Omega_{dm0}=(1-\Omega_{r0}-\Omega_{b0}-\Omega_{dm0})/\Omega_{dm0}$,
and $E(x)\equiv
H(x)/H_0=[\Omega_{r0}e^{-4x}+\Omega_{b0}e^{-3x}+\Omega_{dm0}f_{dm}(x)+(1-\Omega_{r0}-\Omega_{b0}-\Omega_{dm0})f_{de}(x)]^{1/2}.$
Therefore, given the values of the parameters $\Omega_{r0}$,
$\Omega_{b0}$, $\Omega_{dm0}$ and $w$, Eqs.~(\ref{eq8}) and
(\ref{eq9}) can be numerically solved with the initial conditions
$f_{de}(0)=1$ and $f_{dm}(0)=1$. With the resulting functions
$f_{de}(x)$ and $f_{dm}(x)$, we finally obtain the function $E(x)$,
the dimensionless Hubble expansion rate. As aforementioned, in this
work we employ three DE models, namely, the cosmological constant
model ($\Lambda$CDM) with $w=-1$, the constant EOS model (XCDM) with
$w=w_0$, and the time-variable EOS model (CPL) with
$w(x)=w_0+w_1\left(1-e^x\right)$.

To fit the three interacting DE models with observations, we use the
data from the Union2 SNIa (557 data), the BAO from SDSS DR7, the CMB
from 7-year WMAP, the Hubble expansion rate (15 data), and the X-ray
gas mass fraction (42 data). The best-fitted parameters are obtained
by minimizing the sum \be\label{eq33}
\chi^2=\tilde{\chi}^2_{SN}+\chi^2_{BAO}+\chi^2_{CMB}+\chi^2_{H}+\chi^2_{X-ray}.
\ee We obtain the constraints by using a MCMC method.

%=====================================SNIa==========================================

{\it Supernovae.}--- We use the data points of the 557 Union2 SNIa
compiled in Ref.~\cite{14Amanullah:2010vv}. The theoretical distance
modulus is defined as \be\label{eq11} \mu_{th}(z_i)\equiv5\log_{10}
D_L(z_i)+\mu_0, \ee where $z={1/a}-1$ is the redshift,
$\mu_0\equiv42.38-5\log_{10} h$ with $h$ the Hubble constant $H_0$
in units of 100 km/s/Mpc, and the Hubble-free luminosity distance
\be\label{eq12} D_L(z)=(1+z)\int_0^z \frac{dz'}{E(z';{\bm \theta})},
\ee where ${\bm\theta}$ denotes the model parameters.
Correspondingly, the $\chi^2$ function for the 557 Union2 SNIa data
is given by \be\label{eq13}
\chi^2_{SN}({\bm\theta})=\sum\limits_{i=1}^{557}\frac{\left[\mu_{obs}(z_i)-\mu_{th}(z_i)\right]^2}{\sigma^2(z_i)},
\ee where $\sigma$ is the corresponding $1\sigma$ error of distance
modulus for each supernova. The parameter $\mu_0$ is a nuisance
parameter and one can expand Eq.~(\ref{eq13}) as \be\label{eq14}
\chi^2_{SN}({\bm\theta})=A({\bm\theta})-2\mu_0
B({\bm\theta})+\mu_0^2 C, \ee where $A({\bm\theta})$,
$B({\bm\theta})$ and $C$ are defined in Ref.~\cite{Nesseris:2005ur}.
%$$A({\bm\theta})=\sum\limits_{i=1}^{557}\frac{\left[\mu_{obs}(z_i)-\mu_{th}(z_i;\mu_0=0,{\bm\theta})\right]^2}{\sigma_{\mu_{obs}}^2(z_i)},$$
%$$B({\bm\theta})=\sum\limits_{i=1}^{557}\frac{\mu_{obs}(z_i)-\mu_{th}(z_i;\mu_0=0,{\bm\theta})}{\sigma_{\mu_{obs}}^2(z_i)},$$
%$$C=\sum\limits_{i=1}^{557}\frac{1}{\sigma_{\mu_{obs}}^2(z_i)}.$$
Evidently, Eq.~(\ref{eq14}) has a minimum for $\mu_0=B/C$ at
\be\label{eq15}
\tilde{\chi}^2_{SN}({\bm\theta})=A({\bm\theta})-\frac{B({\bm\theta})^2}{C}.
\ee Since $\chi^2_{SN,\,min}=\tilde{\chi}^2_{SN,\,min}$, instead
minimizing $\chi_{SN}^2$ we will minimize $\tilde{\chi}^2_{SN}$
which is independent of the nuisance parameter $\mu_0$.

%=====================================BAO==========================================

{\it Baryon acoustic oscillations.}--- We use the BAO data from SDSS
DR7~\cite{15Percival:2009xn}. The distance ratio ($d_z$) at $z=0.2$
and $z=0.35$ are \be\label{eq16}
d_{0.2}=\frac{r_s(z_d)}{D_V(0.2)},~~
d_{0.35}=\frac{r_s(z_d)}{D_V(0.35)}, \ee where $r_s(z_d)$ is the
comoving sound horizon at the baryon drag
epoch~\cite{16Eisenstein:1997ik}, and \be\label{eq17}
D_V(z)=\left[\left(\int_0^z\frac{dz'}{H(z')}\right)^2\frac{z}{H(z)}\right]^{1/3}
\ee encodes the visual distortion of a spherical object due to the
non Euclidianity of a FRW spacetime. The inverse covariance matrix
of BAO is \bq\label{eq18} (C^{-1}_{BAO}) & = &
\left(\begin{array}{ccc}
30124 & -17227 \\
-17227 & 86977\end{array}\right).\eq The $\chi^2$ function of the
BAO data is constructed as: \be\label{eq19}
\chi_{BAO}^2=(d_i^{th}-d_i^{obs})(C_{BAO}^{-1})_{ij}(d_j^{th}-d_j^{obs}),
\ee where $d_i=(d_{0.2}, d_{0.35})$ is a vector, and the BAO data we
use are $d_{0.2}=0.1905$ and $d_{0.35}=0.1097$.

%=====================================CMB==========================================

{\it Cosmic microwave background.}--- We employ the ``WMAP distance
priors'' given by the 7-year WMAP observations~\cite{17WMAP7}. This
includes the ``acoustic scale'' $l_A$, the ``shift parameter'' $R$,
and the redshift of the decoupling epoch of photons $z_*$. The
acoustic scale $l_A$ describes the distance ratio
$D_A(z_*)/r_s(z_*)$, defined as \be\label{eq21} l_A\equiv
(1+z_*){\pi D_A(z_*)\over r_s(z_*)}, \ee where a factor of $(1+z_*)$
arises because $D_A(z_*)$ is the proper angular diameter distance,
whereas $r_s(z_*)$ is the comoving sound horizon at $z_*$. The
fitting formula of $r_s(z)$ is given by \be\label{eq22}
r_s(z)=\frac{1} {\sqrt{3}} \int_0^{1/(1+z)} \frac{da}
{a^2H(a)\sqrt{1+(3\Omega_{b0}/4\Omega_{\gamma0})a}}. \ee In this
paper, we fix $\Omega_{\gamma0}=2.469\times10^{-5}h^{-2}$,
$\Omega_{b0}=0.02246 h^{-2}$, given by the 7-year WMAP
observations~\cite{17WMAP7}, and $\Omega_{r0}=\Omega_{\gamma0}(1 +
0.2271N_{\rm eff})$ with $N_{\rm eff}$ the effective number of
neutrino species (in this paper we take its standard value, 3.04
\cite{17WMAP7}). We use the fitting function of $z_*$ proposed by Hu
and Sugiyama~\cite{18Hu:1995en} \be\label{eq23}
z_*=1048[1+0.00124(\Omega_{b0}
h^2)^{-0.738}][1+g_1(\Omega_{m0}h^2)^{g_2}], \ee where
$\Omega_{m0}=\Omega_{b0}+\Omega_{dm0}$ and \be\label{eq24}
g_1=\frac{0.0783(\Omega_{b0}h^2)^{-0.238}}{1+39.5(\Omega_{b0}h^2)^{0.763}},\quad
g_2=\frac{0.560}{1+21.1(\Omega_{b0}h^2)^{1.81}}. \ee The shift
parameter $R$ is responsible for the distance ratio
$D_A(z_*)/H^{-1}(z_*)$, given by~\cite{19Bond97} \be\label{eq25}
R(z_*)\equiv\sqrt{\Omega_{m0}H_0^2}(1+z_*)D_A(z_*). \ee
Following~\cite{17WMAP7}, we use the prescription for using the WMAP
distance priors. Thus, the $\chi^2$ function for the CMB data is
\be\label{eq26}
\chi_{CMB}^2=(x^{th}_i-x^{obs}_i)(C_{CMB}^{-1})_{ij}(x^{th}_j-x^{obs}_j),
\ee where $x_i=(l_A, R, z_*)$ is a vector, and $(C_{CMB}^{-1})_{ij}$
is the inverse covariance matrix. The 7-year WMAP
observations~\cite{17WMAP7} give the maximum likelihood values:
$l_A(z_*)=302.09$, $R(z_*)=1.725$, and $z_*=1091.3$. The inverse
covariance matrix is also given in~\cite{17WMAP7}: \bq\label{eq27}
(C_{CMB}^{-1})=\left(\begin{array}{ccc}
2.305 & 29.698 & -1.333 \\
29.698& 6825.27 & -113.180 \\
-1.333& -113.180 &  3.414 \\
\end{array}\right). \eq

%=====================================H(z)==========================================

{\it Hubble expansion rate.}--- For the Hubble parameter $H(z)$,
there are 15 observational data available, where twelve of them are
from Ref.~\cite{20Stern:2009ep}. In addition, in
Ref.~\cite{21Gaztanaga:2008xz}, the authors obtain the additional
three data: $H(z=0.24)=79.69\pm2.32$, $H(z=0.34)=83.8\pm2.96$, and
$H(z=0.43)=86.45\pm3.27$ (in units of km s$^{-1}$ Mpc$^{-1}$). The
$\chi^2$ function for the observational Hubble data is
\be\label{eq28} \chi_{H}^2({\bm\theta})=\sum_{i=1}^{15}
\frac{[H_{th}({\bm\theta};z_i)-H_{ obs}(z_i)]^2}{\sigma^2(z_i)}. \ee

%\begin{table*}[htbp]\caption{\label{table1} The observational $H(z)$
%data~\cite{20Stern:2009ep}.}
%\begin{center}
%\begin{tabular}{c|llllllllllll}
%\hline\hline
% $z$ &\ 0 & 0.1 & 0.17 & 0.27 & 0.4 & 0.48 & 0.88 & 0.9 & 1.30 & 1.43 & 1.53 & 1.75  \\ \hline
% $H(z)\ ({\rm km~s^{-1}\,Mpc^{-1})}$ &\ 74.2 & 69 & 83 & 77 & 95 & 97 & 90 & 117 & 168 & 177 & 140 & 202  \\ \hline
% $1 \sigma$ uncertainty &\ $\pm 3.6$ & $\pm 12$ & $\pm 8$ & $\pm 14$ & $\pm 17$ & $\pm 60$ & $\pm 40$
% & $\pm 23$ & $\pm 17$ & $\pm 18$ & $\pm 14$ & $\pm 40$ \\
%\hline%\hline
%\end{tabular}
%\end{center}
%\end{table*}
%=====================================X-ray==========================================

{\it X-ray gas mass fraction.}--- For the $f_{gas}$ data, we use the
Chandra measurements in Ref.~\cite{22Allen:2007ue}. In the framework
of the $\Lambda$CDM reference cosmology, the X-ray gas mass fraction
is presented as~\cite{22Allen:2007ue,23Goncalves:2009kd}
\be\label{eq29} f_{gas}(z)=\frac{KA\gamma{b(z)}}{1+s(z)}
\left(\frac{\Omega_{b0}}{\Omega_{b0}+\Omega_{dm0}F(z)}\right)
\left[\frac{D_A^{\Lambda{CDM}}(z)}{D_A(z)}\right]^{1.5}, \ee where
the effect of the interaction between dark sectors has been
considered, resulting in an additional function
$F(z)=f_{dm}(-\ln(1+z))/(1+z)^3$. The parameters $K$, $\gamma$,
$b(z)$ and $s(z)$ model the abundance of gas in the clusters. We set
these parameters to their respective best-fit values of
Ref.~\cite{22Allen:2007ue}. $A$ is the angular correction factor,
which is caused by the change in angle for the current test model
$\theta_{2500}$ in comparison with that of the reference cosmology
$\theta_{2500}^{\Lambda{CDM}}$: \be\label{eq30}
A=\left(\frac{\theta_{2500}^{\Lambda{CDM}}}{\theta_{2500}}\right)^\eta\approx
\left(\frac{H(z)D_A(z)}{[H(z)D_A(z)]^{\Lambda{CDM}}}\right)^\eta,
\ee here, the index $\eta$ is the slope of the $f_{gas}(r/r_{2500})$
data within the radius $r_{2500}$, with the best-fit average value
$\eta=0.214\pm0.022$~\cite{22Allen:2007ue}. And the proper (not
comoving) angular diameter distance is given by \be\label{eq31}
D_A(z)=\frac{1}{(1+z)}\int_0^z\frac{dz'}{H(z')}. \ee The $\chi^{2}$
function for the $f_{gas}$ data from the 42 galaxy clusters reads
\be\label{eq32} \chi^2_{X-ray}({\bm\theta})=\sum_{i=1}^{42}
\frac{([f_{gas}({\bm\theta};z_{i})]_{th}-[f_{gas}(z_{i})]_{obs})^{2}}{\sigma^{2}(z_{i})}.
\ee It should be pointed out that the $f_{gas}$ data are rather
crucial for the fitting, since they can be used to break the
degeneracy between the parameters from the interaction and the EOS
of CPL model. So, the inclusion of the $f_{gas}$ data in our fitting
is indispensable.

%============================= section 3 ===========================================

\section{Results}\label{sec3}

Now we fit our interacting models with the observations. For the
interacting $\Lambda$CDM, XCDM and CPL models, the parameters are
${\bm\theta}=\{\Omega_{dm0},~b_0,~b_e,~h\}$,
$\{\Omega_{dm0},~w_0,~b_0,~b_e,~h\}$ and
$\{\Omega_{dm0},~w_0,~w_1,~b_0,~b_e,~h\}$, respectively. We use the
MCMC method and finally we obtain the best-fit parameters and the
corresponding $\chi^2_{min}$. The best-fit, $1\sigma$ and $2\sigma$
values of the parameters with $\chi^2_{min}$ of the three
interacting models are all presented in Table~\ref{table2}.

%=============================  Table. 2 ===========================================

\begin{table*}\caption{The fitting results of the parameters with best-fit values as well as
$1\sigma$ and $2\sigma$ errors in the three interacting DE models.}
\begin{center}
\begin{tabular}{cc|   cc    cc   cc}
\hline\hline model parameters & & $\Lambda$CDM  &  & XCDM & & CPL &
\\ \hline
$\Omega_{dm0}$    && $0.2262^{+0.0242 +0.0356}_{-0.0215 -0.0304}$ &
                   & $0.2267^{+0.0273 +0.0388}_{-0.0237 -0.0327}$ &
                   & $0.2271^{+0.0302 +0.0443}_{-0.0280 -0.0360}$ & \\
$w_0$             && N/A &
                   & $-0.9844^{+0.0915 +0.1282}_{-0.0932 -0.1357}$ &
                   & $-0.9768^{+0.3046 +0.4322}_{-0.2090 -0.2635}$ & \\
$w_1$             && N/A &
                   & N/A &
                   & $-0.0455^{+0.9426 +1.2742}_{-1.8125 -2.7810}$ & \\
$b_0$             && $0.0793^{+0.1220 +0.1797}_{-0.1092 -0.1580}$ &
                   & $0.0787^{+0.1353 +0.1962}_{-0.1159 -0.1654}$ &
                   & $0.0799^{+0.1512 +0.2039}_{-0.1342 -0.1894}$ & \\
$b_e$             && $-0.3274^{+0.4646 +0.6543}_{-0.5140 -0.7644}$ &
                   & $-0.3216^{+0.5007 +0.7082}_{-0.5771 -0.8317}$ &
                   & $-0.3290^{+0.5855 +0.8379}_{-0.6447 -0.8989}$ & \\
$h$               && $0.7120^{+0.0154 +0.0224}_{-0.0155 -0.0218}$ &
                   & $0.7094^{+0.0224 +0.0324}_{-0.0219 -0.0315}$ &
                   & $0.7097^{+0.0259 +0.0348}_{-0.0257 -0.0357}$ & \\
\hline $\chi^{2}_{min}$  && 595.968 & & 595.815  & & 595.808  & \\
\hline\hline
\end{tabular}
\label{table2}
\end{center}
\end{table*}

%=======================================Fig=========================================
\begin{figure}[htbp]
 \centering \noindent
 \includegraphics[width=8cm]{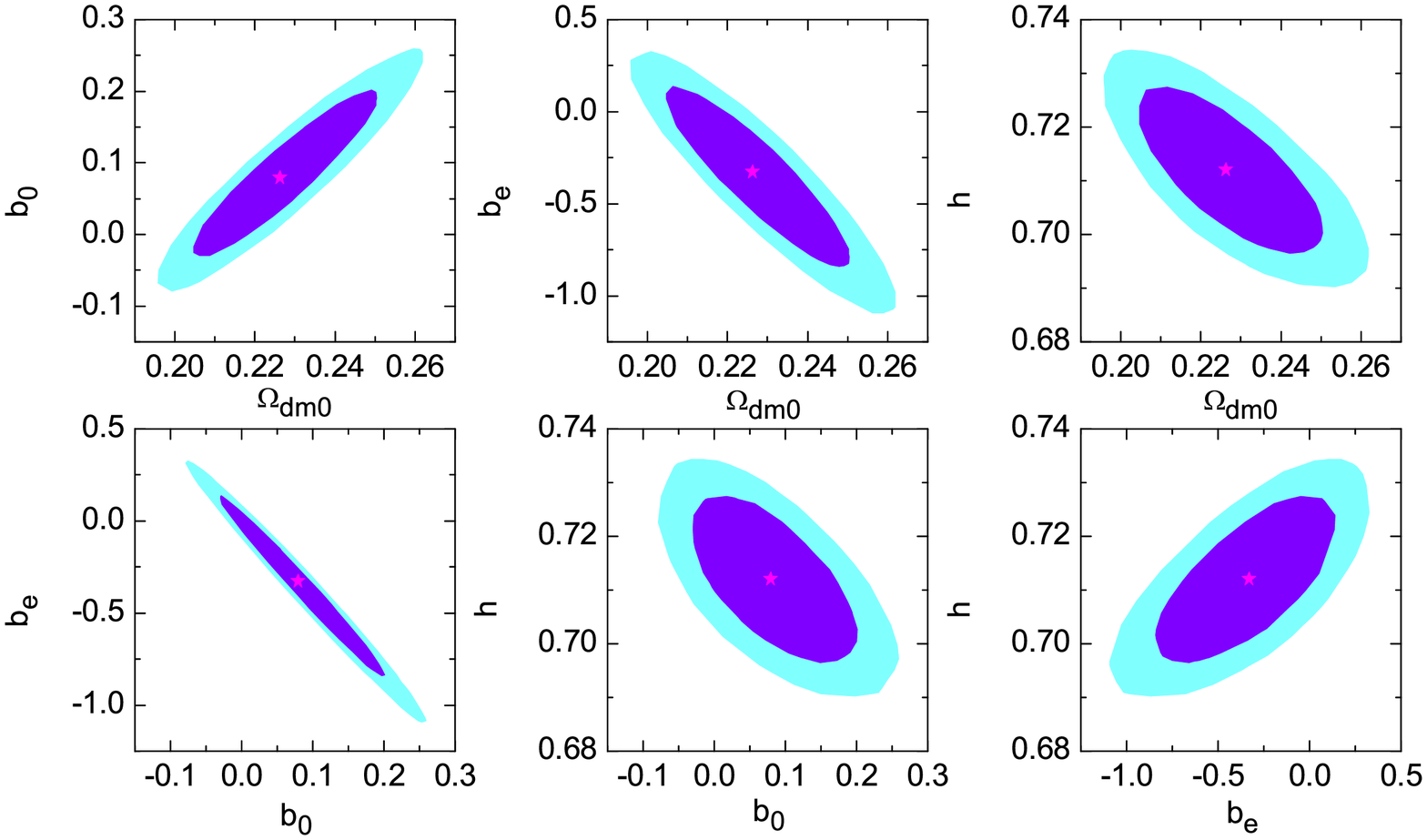}
 \caption{\label{fig1:LCDM}The probability contours at
 $1\sigma$ and $2\sigma$ confidence levels in the parameter planes for the interacting $\Lambda$CDM model.}
 \end{figure}

 \begin{figure*}[htbp]
 \centering \noindent
 \includegraphics[width=14cm]{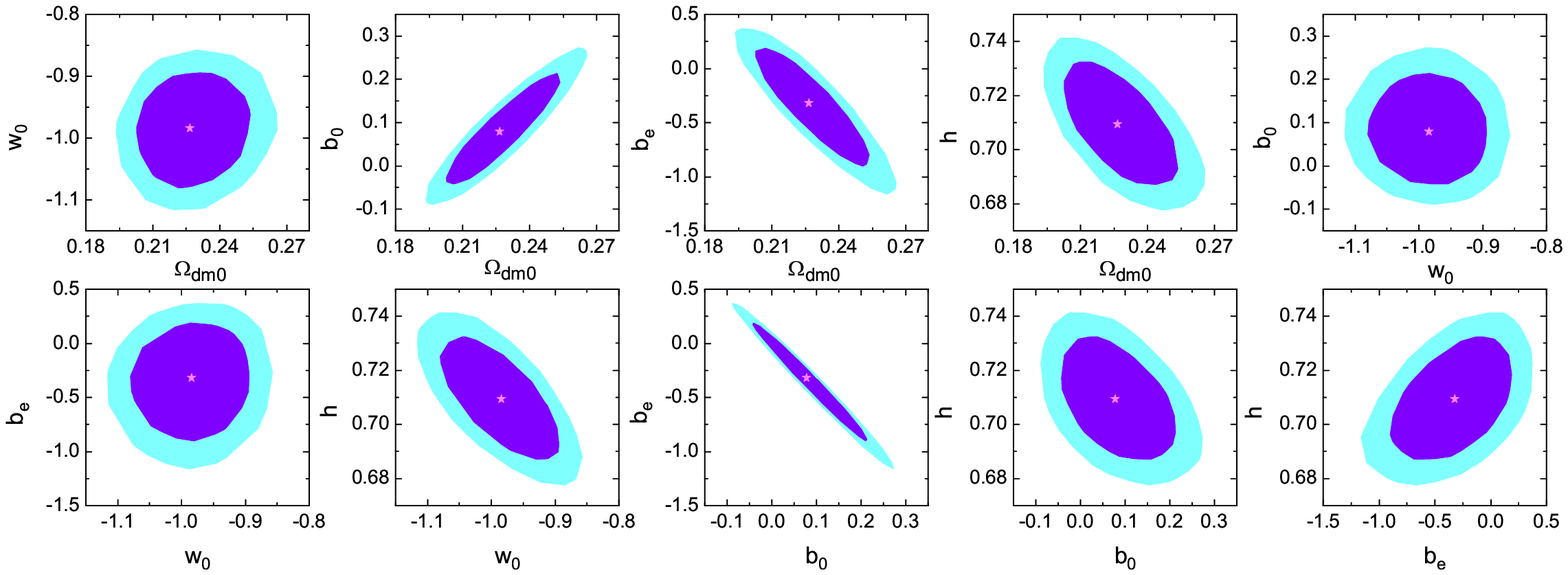}
 \caption{\label{fig2:XCDM}The probability contours at $1\sigma$ and $2\sigma$ confidence levels in the parameter planes for the interacting XCDM model.}
 \end{figure*}

 \begin{figure*}[htbp]
 \centering \noindent
 \includegraphics[width=14cm]{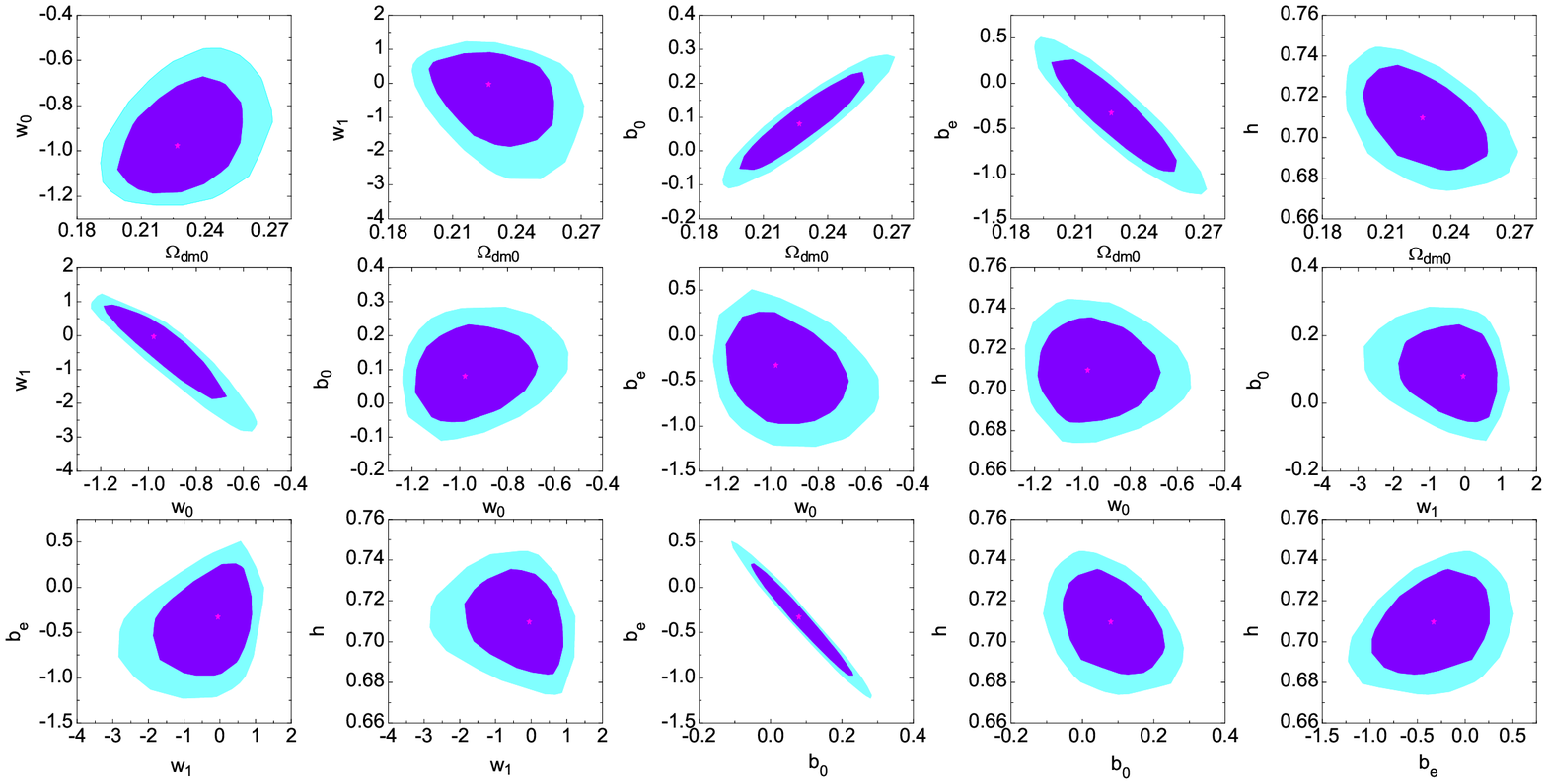}
 \caption{\label{fig3:CPL}The probability contours at $1\sigma$ and $2\sigma$ confidence levels in the parameter planes for the interacting CPL model.}
 \end{figure*}

\begin{figure}[htbp]
 \centering \noindent
 \includegraphics[width=5cm]{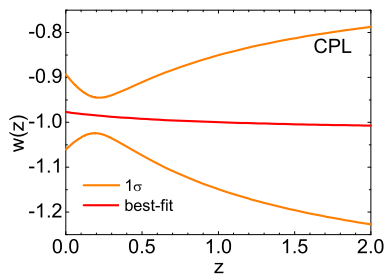}
 \caption{\label{fig4:WCPL}The reconstructed evolutionary history for $w(z)$ of the interacting CPL model.}
 \end{figure}

 \begin{figure*}[htbp]
 \centering\noindent
 \includegraphics[width=5cm]{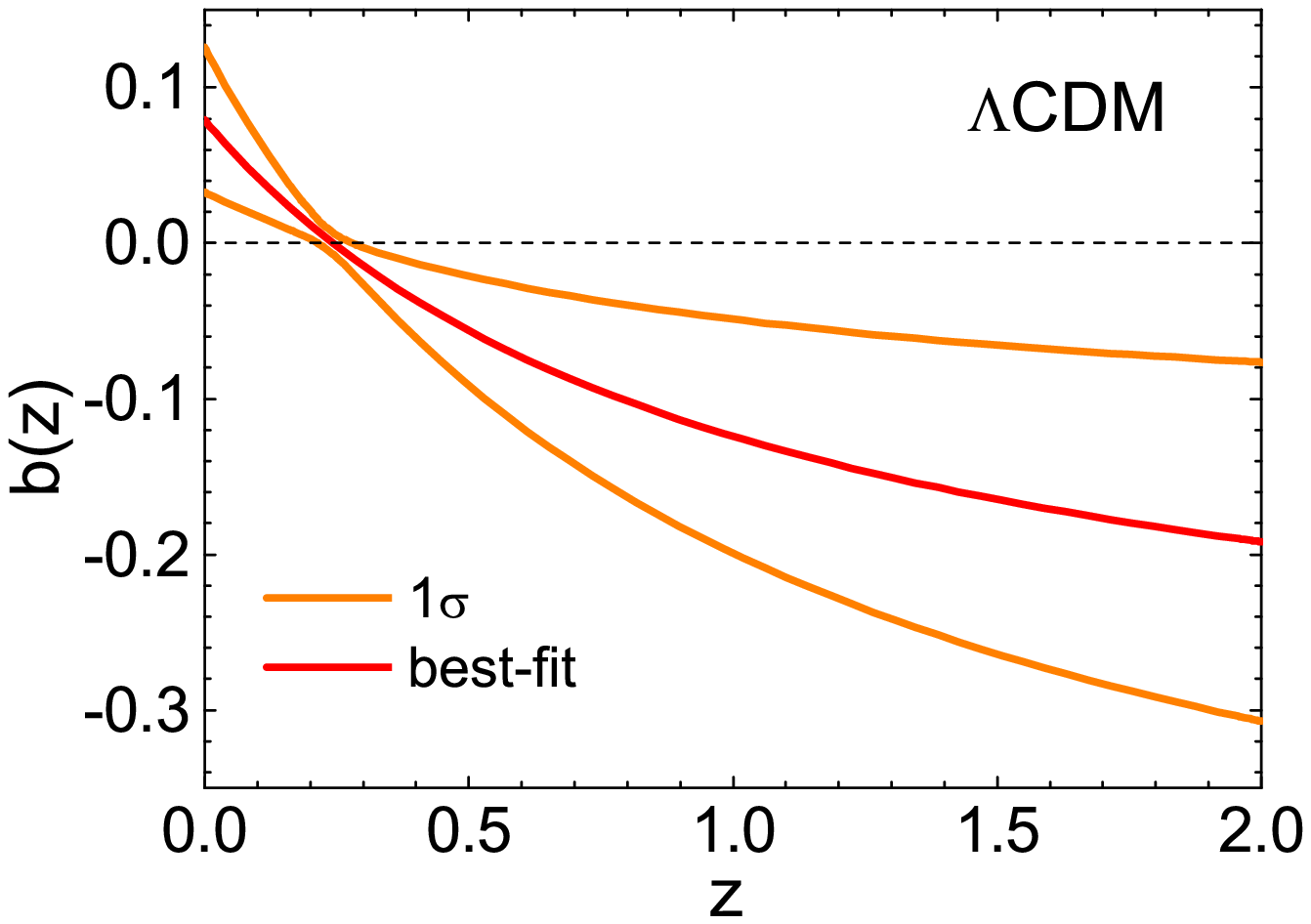}\hskip.1cm
 \includegraphics[width=5cm]{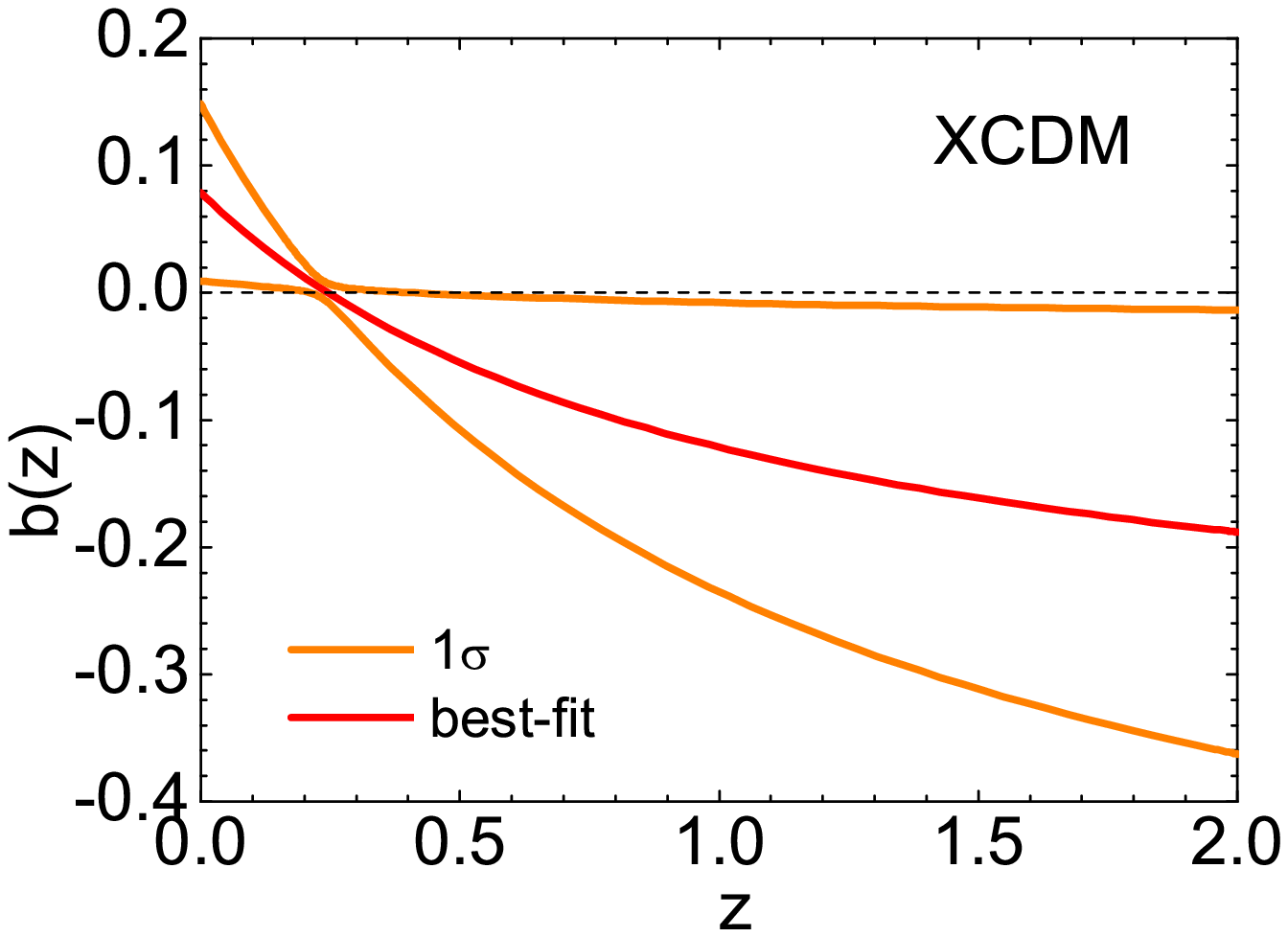}\hskip.1cm
 \includegraphics[width=5cm]{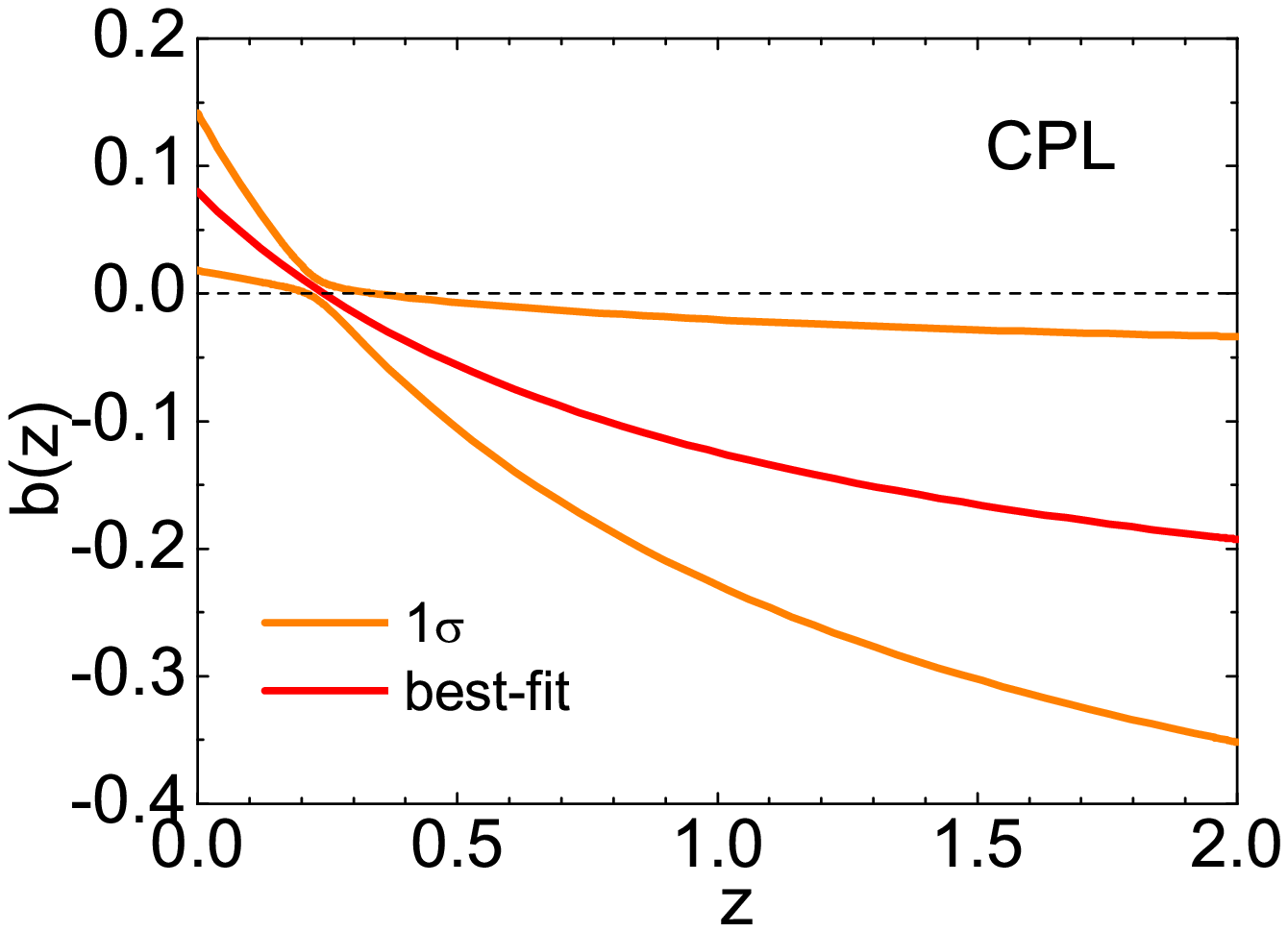}
 \caption{\label{fig5:QALL} The reconstructed evolutionary histories for $b(z)$
 in the three interacting models. The dashed line in each plot represents the noninteracting line.}
 \end{figure*}

%===================================================================================
%===================================================================================

Figure~\ref{fig1:LCDM} shows the likelihood contours for the
interacting $\Lambda$CDM model. For this model, we have
$\Omega_{dm0}=0.2262$, $b_0=0.0793$, $b_e=-0.3274$ and $h=0.7120$,
with $\chi^2_{min}=595.968$. We plot the likelihood contours for the
interacting XCDM model in Fig.~\ref{fig2:XCDM}. For the interacting
XCDM model, the fitting results are $\Omega_{dm0}=0.2267$,
$w_0=-0.9844$, $b_0=0.0787$, $b_e=-0.3216$ and $h=0.7094$, with
$\chi^2_{min}=595.815$. For the interacting CPL model, we obtain the
fitting results: $\Omega_{dm0}=0.2271$, $w_0=-0.9768$, $w_1=-0.0455$
$b_0=0.0799$, $b_e=-0.3290$ and $h=0.7097$, with
$\chi^2_{min}=595.808$. The likelihood contours for this case are
shown in Fig.~\ref{fig3:CPL}.

From the fitting results, we first notice that the EOS of DE is near
$-1$ in both the constant $w$ and time-dependent $w$ scenarios. In
the constant $w$ case, we get the best-fitted value $w_0=-0.9844$,
rather close to $-1$, albeit slightly tends to a quintessence
($w>-1$). For the CPL case, the best-fitted values are $w_0=-0.9768$
and $w_1=-0.0455$; we also notice that $w_0$ is fairly near $-1$ and
$w_1$ is close to 0. So, in the interacting DE model with a running
coupling, even if the EOS of DE is allowed to vary, the data still
favor a slightly evolving EOS with the value approaching $-1$. To
see this clearly, we plot the reconstructed $w(z)$ for the
interacting CPL model in Fig.~\ref{fig4:WCPL} where the best fit and
the 1$\sigma$ uncertainties are shown. According the fitting
results, we realize that a time-varying vacuum scenario is favored
by the data. The time-varying vacuum model has been discussed
extensively \cite{runvac,Odintsov:2005ju}. Our work indicates that
the scenario of a time-varying vacuum with a running coupling
deserves more further investigations.

For the running coupling $b(z)$, we obtain similar results in all
the three scenarios. From Figs.~\ref{fig1:LCDM}--\ref{fig3:CPL} we
see that the parameters $b_0$ and $b_e$ are in strong
anti-correlation. The best-fitted values for $b_0$ and $b_e$ are:
$b_0\approx 0.08$ and $b_e\approx-0.3$, implying that the coupling
$b(z)$ crosses the noninteracting line $b=0$ during the cosmological
evolution, and the sign changes from negative to positive. Such a
feature for the interaction is favored by the data at about
1$\sigma$ level. To see the crossing phenomenon clearly, we
reconstruct the evolution of the coupling $b(z)$ by using a Fisher
Matrix technique, shown in Fig.~\ref{fig5:QALL}. From this figure we
read out that the crossing happens at $z=0.2-0.3$. Our results tell
such a story: at early times when DM dominates the universe, the
energy transfer direction is from DM to DE, and at late times when
DE becomes dominant, the decay direction reverses, from DE to DM.
The above phenomenon is favored at 1$\sigma$ CL. Note that once the
reconstruction of $b(z)$ is performed based on a Monte Carlo method,
the fluctuations will become larger, but the above distinctive
feature still stands by at about 1$\sigma$ CL.

In the work of Cai and Su~\cite{12Cai:2009ht}, a redshift-binned
parametrization method is used, but the limitation is that it is
hard to go beyond two parameters for tight constraints. Also, in
Ref.~\cite{12Cai:2009ht}, as the data give only weak constraint for
$z>1.8$, the additional assumption $Q=0$ for $1.8<z<1090$ is made.
In our method, the parametrization $Q(a)=3b(a)H_0\rho_0$ with
$b(a)=b_0a+b_e(1-a)$ has only two parameters, and the early-time
interaction can also be described. Our fitting results support the
conclusion of Cai and Su~\cite{12Cai:2009ht} that the interaction
$Q(z)$ crosses the noninteracting line, in a distinct way. The
limitations of our parametrization are: (1) whether or not there is
some oscillation feature in the interaction cannot be read out, and
(2) the future evolution cannot be described owing to the fact that
$b(z)$ will diverge as $z$ approaches $-1$, so the predictive power
of this parametrization is lost (see \cite{Ma:2011nc} for a similar
case for $w(z)$). We will go beyond these limitations in the future
work. In addition, now that the analysis of the current
observational data provides a hint that the interaction between dark
sectors might change sign during the cosmological evolution, more
general phenomenological forms for the interaction describing the
sign-changeable or oscillatory feature should be seriously
considered.

%============================= section 4 ===========================================

\section{Conclusion}\label{sec4}

Since the knowledge about the micro-origin of dark sector
interaction is absent, one has no way to know the form of the
interaction term $Q$ from a microscopic theory. A popular way to
investigate the interacting dark energy is to assume a specific
phenomenological form for $Q$; for instance, $Q\propto H\rho$ or
$Q\propto \rho$, where $\rho$ denotes the energy density of dark
sectors. However, the choice of the phenomenological forms is rather
arbitrary. In the face of this situation, let us recall the method
used to probe the dynamical evolution of the EOS of DE, $w(z)$, with
the observational data. Owing to our ignorance of DE, for probing
the dynamical evolution of $w$, one has to parameterize $w$
empirically, usually using two parameters, e.g., the most widely
used CPL form, $w(a)=w_0+w_1(1-a)$. Inspired by this method, we try
to parameterize the evolution of $Q$ in a similar way, and then use
the observational data to probe the dynamical evolution of $Q$. Such
a way may provide a guidance for finding a reasonable
phenomenological form for $Q$.

In this paper, we have put forward a running coupling scenario for
describing the interaction between DE and DM. In this scenario, the
dark sector interaction has the form $Q(a)=3b(a)H_0\rho_0$, where
$b(a)$ is the coupling which is variable during the cosmological
evolution. We have proposed a parametrization form for the running
coupling: $b(a)=b_0a+b_e(1-a)$. So, at the early times the coupling
is given by a constant $b_e$, while today the coupling is described
by another constant, $b_0$. We have constrained the parameters $b_0$
and $b_e$ with the observational data currently available, including
the Union2 SNIa, BAO (from SDSS DR7), CMB (from 7-year WMAP),
$H(z)$, and X-ray gas mass fraction data. It should be stressed that
the $f_{gas}$ data are very crucial in our analysis since they are
helpful in breaking the degeneracy between the parameters. And, in
our analysis, we employed three DE model, namely, the $\Lambda$CDM,
XCDM and CPL models.

Our fitting results show that the EOS of DE, $w$, is close to $-1$
both in the constant $w$ (XCDM) and time-dependent $w$ (CPL) cases.
Thus, a time-varying vacuum scenario is favored by the data,
according to this analysis. In addition, for the running coupling
$b(z)$, the results are also similar in all the three DE models. The
parameters $b_0$ and $b_e$ are in strong anti-correlation, and the
best-fitted values are: $b_0\approx 0.08$ and $b_e\approx-0.3$. This
implies that the coupling $b(z)$ crosses the noninteracting line
($b=0$) and the sign changes from $b<0$ to $b>0$. The reconstruction
of $b(z)$ indicates that the crossing of the noninteracting line
happens at around $z=0.2-0.3$, and the crossing behavior is favored
at about 1$\sigma$ CL. Therefore, our work tells a story about the
interacting DE model: DE is a time-varying vacuum; the coupling
between DE and DM runs with the expansion of the universe; at early
times when DM dominates the universe, DM decays to DE, while at late
times when DE becomes dominant, DE begins to decay to DM. If the
above scenario is true, then we should pay more attention to the
time-varying vacuum model, and seriously consider how to construct a
sign-changeable interaction between dark sectors phenomenologically.
For the time-varying vacuum model, previous work neglects the
perturbations of DE and only treats the decaying vacuum as a
background, however, if DE interacts with DM, the perturbations of
DE should also be taken into account even if $w=-1$. This deserves
further investigations.

Finally, we discuss the limitations of our parametrization for the
running coupling. Our parametrization cannot describe the possible
oscillation in the dark sector interaction. Moreover, such a
parametrization form cannot predict the future evolution of the
interaction, since the coupling $b(z)$ will encounter divergency in
the far future. We will go beyond these limitations in our future
work.

%============================= acknowledgments =====================================

\begin{acknowledgments}
This work was supported by the Natural Science Foundation of China
under Grant Nos.~10705041 and 10975032, as well as the National
Innovation Experiment Program for University Students.
\end{acknowledgments}

%============================= references ==========================================

\end{document}